\documentclass{PoS}

\title{Calculation of Neutron EDM in quenched and full QCD}
 
\ShortTitle{Calculation of Neutron EDM in quenched and full QCD}

\author{CP-PACS Collaboration: 
   \speaker{E. Shintani,$^1$\thanks{E-mail:shintani@het.ph.tsukuba.ac.jp}}
   S. Aoki,$^{1,2}$ N. Ishizuka,$^{1,3}$ K. Kanaya,$^1$
   Y. Kuramashi,$^{1,3}$ M. Okawa,$^4$ A. Ukawa,$^{1,3}$ T. Yoshi\'e,$^{1,3}$.\\
\llap{$^1$} Graduate School of Pure and Applied Sciences, University of Tsukuba\\
\llap{$^2$} Riken BNL Research Center, Brookhaven National Laboratory\\
\llap{$^3$} Center for Computational Sciences, University of Tsukuba\\
\llap{$^4$} Department of Physics, Hiroshima University
}

\abstract{
We report on a direct lattice calculation of the neutron EDM(NEDM) 
using the external electric field method in both quenched and full QCD. 
In quenched QCD, we use a $24^3\times 32$ lattice at $\beta=2.6$ with
the Iwasaki gauge action and the clover fermion action to examine the 
viability of this method. In particular we investigate
possible effects of violation of the periodic boundary condition 
of  the external electric field on the NEDM signal.
We also study the quark mass dependence of NEDM in quenched QCD, and
observe that NEDM seems to remain non-zero toward the chiral limit
because of the quenched artifact. 
In 2-flavor full QCD we employ configurations generated by the CP-PACS
collaboration on a $24^3\times 48$ lattice at $\beta=2.1$ with the same
gluon and quark actions as in the quenched case.
Since the number of configurations is limited, we employ 8 different
source points per one configuration and take an average over them.
Our preliminary result at three quark masses 
($m_{PS}/m_V\simeq 0.81,\,0.76,\,0.69$) indicates that
non-zero value for NEDM can be obtained in full QCD.
Statistical errors, however, are still too large to show the theoretically
expected behaviour for NEDM in full QCD that it vanishes in the chiral limit.
\vspace{0.7cm}\\  UTCCS-P-26}

\FullConference{XXIV International Symposium on Lattice Field Theory\\
                 July 23-28 2006\\
                 Tucson Arizona, US}
 
\begin{document}
\section{Introduction}
Calculating the neutron electric
dipole moment (NEDM) in the presence of the $\theta$ term of QCD has been 
known to be a difficult task.
Various effective models in fact give order of magnitude different
results for NEDM.  
Nonetheless, combining these rough estimates with the experimental upper bound on
NEDM given by $d_N\le 6.3\times 10^{-13}{\rm e\cdot fm}$ \cite{Harris}
yields a bound on $\theta$, $\theta< 10^{-10}$, which confirms
the existence of the strong CP problem.
Our aim here is to give a more accurate and reliable estimate of NEDM
in the presence of the $\theta$ term based on lattice QCD calculations. 
Such a result will be indispensable in order to extract the value of $\theta$
from a non-zero value of NEDM should it be observed in future experiments. 

The first lattice calculation of NEDM failed to obtain a signal\cite{Aoki}. 
Fifteen years later we proposed a formulation to extract NEDM form factors and 
numerically verified that the method works in quenched QCD with the
domain-wall fermion\cite{Shintani}. 
In the same year, this formulation was employed in a full QCD calculation  
with the domain-wall fermion\cite{RBC}. 
They, however, could obtain only an upper bound on the NEDM value. 
At lattice 2005, we returned to the original external electric
field method proposed in Ref.~\cite{Aoki}, and reported 
our preliminary results for NEDM from this method with both domain-wall and clover
fermions in quenched QCD\cite{Shintani2}.

In this report we carry out further numerical studies in the external 
electric field method with the clover fermion. 
We first investigate the problem with this method that a large
discontinuity (gap) appears in the uniform electric field at the
boundary in the time direction violating the periodic
boundary condition. From this investigation we learn how we can minimize
the unwanted effect of the gap to NEDM signals. 
Using this knowledge, we apply the method to both quenched and full 
QCD calculations at several quark mass parameters
 in order to investigate the quark mass dependence of NEDM.

\section{Formulation}
In our calculation, NEDM is extracted from the interaction term between
the electric field $\vec E$ and the neutron spin $\vec S$ in the Hamiltonian: 
\begin{equation}\label{eq:hamil}
  \delta H_{\rm CP} = d_N(\theta) \vec S\cdot\vec E.
\end{equation}
where $d_N(\theta)$ represents the magnitude of NEDM, and $\theta$ is
the CP violating phase of the strong interaction. For small $\theta$
we can expand $d_N(\theta) = d_N\theta + O(\theta^3)$, 
and hereafter we ignore higher order terms in $\theta$.

A uniform and static electric field in Minkovski space is introduced by
replacing the link variable as 
\begin{equation}
  U_i(x) \longrightarrow \tilde U_i(x) = U_i(x)e^{q_eE_it},\quad  
  U_i^{\dag}(x) \longrightarrow \tilde U_i^\dag(x) = U^\dag_i(x)e^{-q_eE_it},
\end{equation}
where $q_e$ represents the quark charge: $q_e=2/3$ for up quark and $q_e=-1/3$ 
for down quark. 
This replacement, however, destroys the periodicity of $U_i(x)$
at the boundary $t=T$:
\begin{equation}
  \tilde U_i(\vec x,t=T) = U_i(\vec x,T)e^{q_eE_iT}\ne\tilde U_i(\vec x,t=0) = U_i(\vec x,t=0),
\end{equation}
As a consequence, uniformity of the electric field is violated between
$t=0$ and $t=T$. In section~\ref{sec:result_quench} we will numerically 
investigate the effect of this non-uniformity to the EDM signal.
 
To include the effect of the $\theta$ term,
we reweight the nucleon propagator with $e^{i\theta Q}$:
\begin{equation}
  \langle N_{\alpha}\bar N_{\beta}\rangle_\theta = 
  \left[ \langle N\bar Ne^{i\theta Q}\rangle\right]_{\alpha\beta}
\end{equation}
where $\alpha,\beta$ denotes spinor indicies of the nucleon.
We may reduce the sign problem due to the complex reweighting factor
$e^{i\theta Q}$ by taking $\theta$ small. On the other
hand, we have to accumulate a large number of configurations
to ensure correct distribution of topological charge $Q$, since 
results of NEDM strongly depend it.

For the electric field in $z$ direction, $E=(0,0,E)$,  
the ratio of nucleon propagator between spin up and down components has
the following $t$ dependence:
\begin{equation}
  R^{\rm naive}_3(E,t;\theta) = \frac{\langle N_1\bar N_1\rangle_\theta(E,t)}
                                     {\langle N_2\bar N_2\rangle_\theta(E,t)}
  = Z_N(E,\theta)e^{-d_N\theta Et} + \cdots ,
\end{equation}
where $Z_N(E,\theta)$ is constant in $t$ but depends on $E$ and $\theta$, and 
the dots represent contributions from excited states.
We further take the following triple ratio in order to subtract fake
contributions from $E=0$ and $\theta=0$ caused by finite statistics.
\begin{equation}
R_3(E,t;\theta) = \frac{R_3^{\rm naive}(E,t;\theta)}
{R_3^{\rm naive}(0,t;\theta)}
\times
\frac{R_3^{\rm naive}(0,t;\theta=0)}{R_3^{\rm naive}(E,t;\theta=0)}.
\end{equation}
We finally take the ratio between $E$ and $- E$, in order to remove
$O(E^2)$ contributions as
\begin{equation}
  R_3^{\rm corr}(E,t;\theta) = \frac{R_3(E,t;\theta)}{R_3(-E,t;\theta)}
  = \frac{R_3^{\rm naive}(E,t;\theta)}{R_3^{\rm naive}(-E,t;\theta)}
    \frac{R_3^{\rm naive}(E,t;\theta=0)}{R_3^{\rm naive}(-E,t;\theta=0)}
  = Z_Ne^{-2d_N\theta Et} + \cdots .
\label{eq:R3corr}
\end{equation}
In section~\ref{sec:results} 
we will show the time dependence of $R_3(E,t;\theta)$ for both $E$ and
$-E$, and we will also plot the effective mass of $R_3^{\rm corr}$.

\section{Numerical results}\label{sec:results}
\begin{figure}
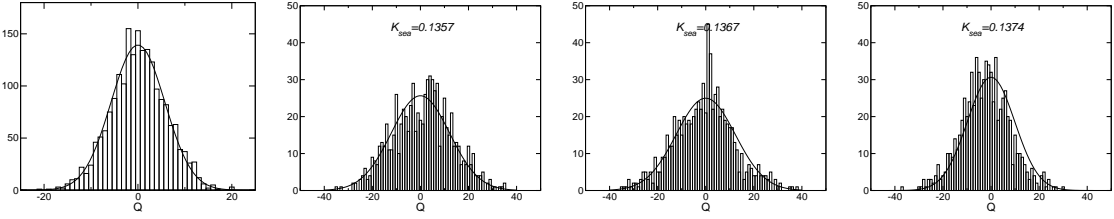

\begin{center}
\vskip -3mm
\includegraphics[width=34mm, angle=0] {Fig/hist.Qeunch.cool20.eps}
\hspace{.2cm}
\includegraphics[width=34mm, angle=0] {Fig/hist.cool50.K01357.eps}
\hspace{.2cm}
\includegraphics[width=34mm, angle=0] {Fig/hist.cool50.K01367.eps}
\hspace{.2cm}
\includegraphics[width=34mm, angle=0] {Fig/hist.cool50.K01374.eps}
\vskip -2mm
\caption{Histogram of topological charge distribution in quenched 
QCD at $\beta=2.6$ (first) and full QCD at $\beta=2.1$ with
$K_{\rm sea}=0.1357$ (second), $K_{\rm sea}=0.1367$ (third), $K_{\rm sea}=0.1374$ (fourth).}
\vskip -5mm
\label{fig:hist}
\end{center}
\end{figure}

\begin{figure}
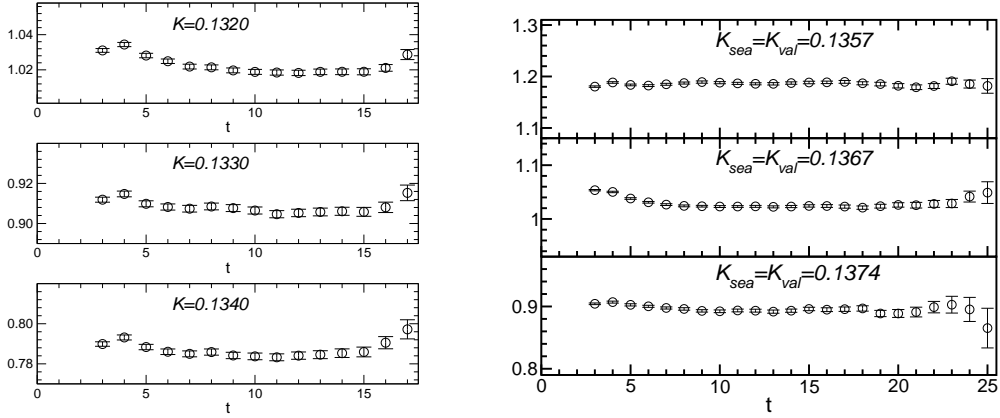

\begin{center}
\vskip 2mm
\includegraphics[width=55mm, angle=0] {Fig/NEmass.Ns24clv.eps}
\hspace{1cm}
\includegraphics[width=65mm, angle=0] {Fig/Nmass.eps}
\vskip -2mm
\caption{Effective mass plot of nucleons in quenched QCD(left) and 
full QCD(right).}
\label{fig:NEmass}
\vskip -5mm
\end{center}
\end{figure}

In our calculation we employ the RG improved (Iwasaki) gauge action and the 
clover fermion action with the tadpole improved tree value of $c_{sw}$.
The $O(a^2)$ improved field theoretical definition 
of the topological charge is employed, after 20 cooling steps 
in quenched QCD and 50 cooling steps in full QCD.
In Fig.~\ref{fig:hist} the histogram of the topological charge
distribution is plotted in quenched QCD at $\beta=2.6$ (leftmost)
and in full QCD at $\beta=2.1$ and $K_{\rm sea}=0.1357$ (second from left), 
$K_{\rm sea}=0.1367$ (third), $K_{\rm sea}=0.1374$ (fourth).
Smeared sources with the form of $Ae^{-Br}$ is employed in our
calculation. In Fig.~\ref{fig:NEmass}, the effective mass of nucleon at 
$E=0$ and $\theta=0$ is plotted in quenched QCD (left) and full QCD (right)
at several values of the quark mass.


\subsection{Effects of non-uniform electric field in quenched QCD}
\label{sec:result_quench}

\begin{figure}
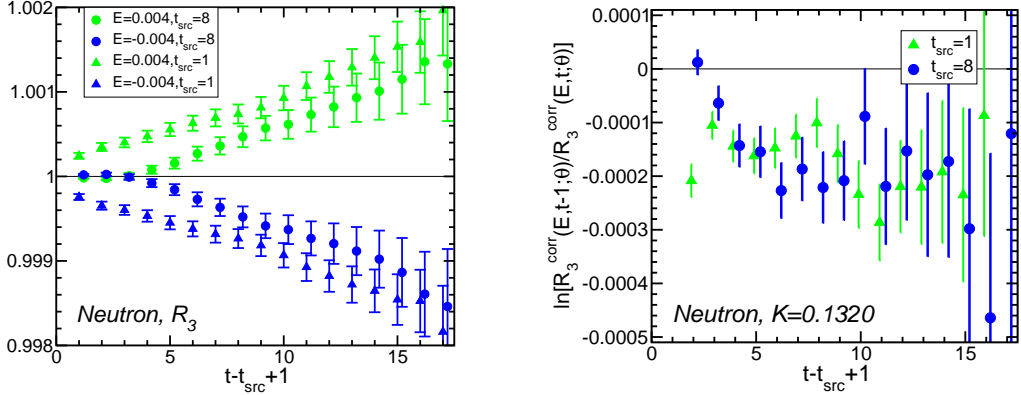

\begin{center}
\includegraphics[width=60mm, angle=0] {Fig/NEdn.E0004.theta01.Ns24clv.z.SRC8.corr.eps}
\hspace{1cm}
\includegraphics[width=63mm, angle=0] {Fig/NeffEdn.E0004.theta01.Ns24clvK01320.SRC8.eps}
\vskip -2mm
\caption{Time dependence of $R_3(E,t;\theta)$ (left) and
effective mass (right) for $t_{\rm src}=1$ and $t_{\rm src}=8$ in
quenched QCD.}
\label{fig:NEdn_Quench_src}
\vskip -5mm
\end{center}
\end{figure}

We first investigate the effect of a non-uniform electric field at the
boundary to EDM signals, using quenched QCD on a $24^3\times 32$ lattice
at $\beta=2.6$, which corresponds to $a^{-1}\simeq 2$ GeV. 
We use $E=0.004$ and $\theta=0.1$ in this study.
We always put the gap (discontinuity) of the electric field between  
$t=T$ and $t=1$, denoting this as $t_E=1$.
For comparison,
we place the nucleon source at two different time slices: $t_{\rm src}=1$ and
$t_{\rm src}=8$.
To reduce the computational cost we take a heavy fermion mass 
$K=0.1320$ corresponding to $m_{PS}/m_V\simeq 0.85$.
On the left of Fig.~\ref{fig:NEdn_Quench_src} we compare the time
dependence of $R_3(E,t;\theta)$ between $t_{\rm src}=1$ and $t_{\rm src}=8$.
We observe that the $t$ dependences are much different between the two cases
near source points ($t-t_{\rm src}+1\le 5$). On the other hand, 
at large $t$ ($t-t_{\rm src}+1\ge 8$), the two cases show similar 
behaviors. This indicates that the gap affects the EDM signal near
$t=t_E$ while its influence almost disappears at large $t$ which satisfies
$t - t_E \ge 8$. Therefore the result with $t_{\rm src}=8$ is expected
to be free from the influence of the gap. 

This tendency is also seen from the comparison of the effective mass
plot between two sources, given on the right of 
Fig.~\ref{fig:NEdn_Quench_src}. 
We observe a plateau at $6\le t-t_{\rm src}+1\le 10$ for the case of 
$t_{\rm src}=8$. For the case of $t_{\rm src}=1$, 
there is a candidate for a plateau at $6\le t-t_{\rm src}+1\le 10$. 
However, the value in this range of $t$ disagrees with that from the 
plateau of $t_{\rm src}=8$ data.
Therefore, we should conclude that this is not a real plateau, but 
a fake effect due to the gap.  In fact, looking further away from the 
source, we observe a plateau at the range 
$8\le t-t_{\rm src}+1\le 12$ for $t_{\rm src}=1$ whose value is in 
agreement with that for $t_{\rm src}=8$.

Our study concludes that we should place the source point as far from
the gap as possible in order to avoid the influence of the gap to EDM signals.
The condition that $t_{\rm src}- t_E\ge 8$ seems necessary in this
method for the EDM calculation.

\subsection{Preliminary results in full QCD}

We apply our method of EDM calculations to $N_f=2$ flavor full QCD 
configurations on a $24^3\times 48$ lattice at $\beta=2.1$ 
corresponding to $a^{-1}\simeq 2$ GeV, 
generated by CP-PACS collaboration\cite{CP-PACS} using
RG improved gauge and tadpole improved clover fermion actions. 
In this calculation we take $E=0.004$ and $\theta=0.025$ and employ four
values for valence quark masses, which are equal to sea quark masses of
full QCD ( $K_{\rm sea}=0.1357$, $0.1367$, $0.1374$, $0.1382$)
in order to perform the chiral extrapolation.

As shown in section~\ref{sec:result_quench}, the influence of the
non-uniform electric field  at the boundary to the EDM signal is
largely suppressed if $t_{\rm src}-t_E\ge 8$.
In this full QCD calculation, we take the maximal separation, namely
$t_{\rm src}-t_E = 24$ on a $24^3\times 48$ lattice.
Since the distance between the gap and the source for the positive time
direction ($t_{\rm src}-t_E=24$) is same as the one for the negative
time direction ($48+t_E -t_{\rm src}=24$) in this setup due to the
periodic boundary condition in time, we can utilize both 
forward and backward propagations of nucleon for EDM calculations.

Our study in quenched QCD indicates that the number of full QCD configurations
available (about 770) is not sufficient for our method of 
EDM calculations.
In order to effectively increase statistics, 
we employ 8 different source points per one configuration,
choosing $(t_{\rm sec},t_E)=$(6,30), (12,36), (18,42), (24,48), (30,6), (36,12), (42,18)
, (48,24).  The decrease of statistical errors we observe in our results 
indicates that data from these source points are almost independent.
Therefore the total number of statistics is more than 12,000, 
including forward and backward propagations.

\begin{figure}
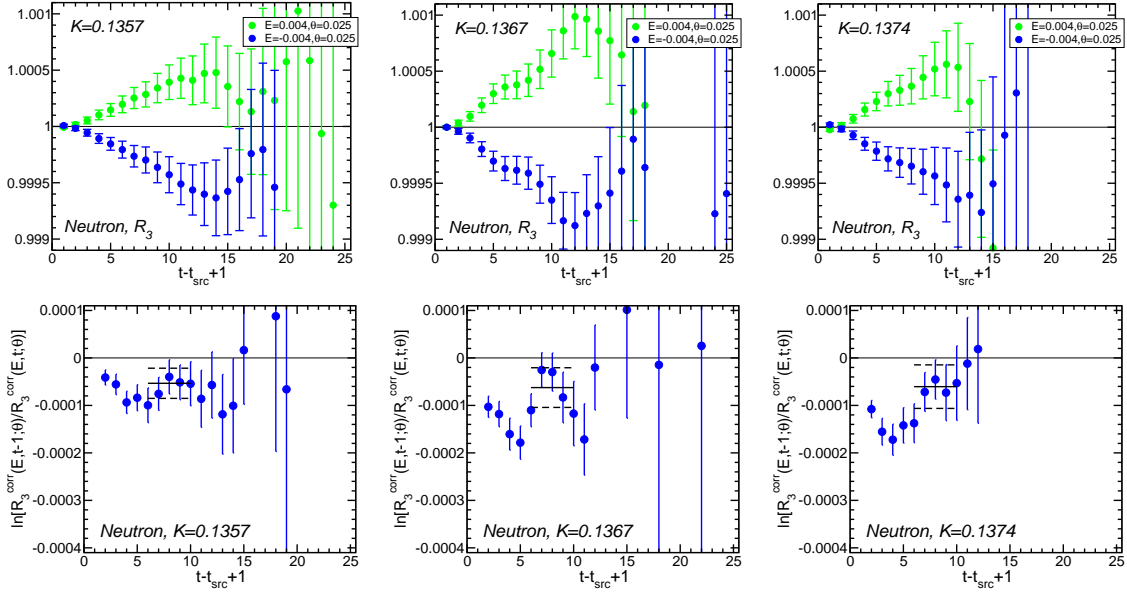

\begin{center}
\vskip -3mm
\includegraphics[width=47mm, angle=0] {Fig/NEdn.K01357.E0004.theta0025.corr.eps}
\hspace{.2cm}
\includegraphics[width=47mm, angle=0] {Fig/NEdn.K01367.E0004.theta0025.corr.eps}
\hspace{.2cm}
\includegraphics[width=47mm, angle=0] {Fig/NEdn.K01374.E0004.theta0025.corr.eps}\\
\vskip .2cm
\includegraphics[width=47mm, angle=0] {Fig/NeffEdn.E0004.theta0025.K01357.eps}
\hspace{.2cm}
\includegraphics[width=47mm, angle=0] {Fig/NeffEdn.E0004.theta0025.K01367.eps}
\hspace{.2cm}
\includegraphics[width=47mm, angle=0] {Fig/NeffEdn.E0004.theta0025.K01374.eps}
\vskip -2mm
\caption{Time dependence of $R_3$ function(top) and effective mass plot
of $R_3^{\rm corr}$(bottom) at $K=K_{\rm sea}=K_{\rm val}=0.1357,\,0.1367,\,0.1374$ 
averaged over 8 source sets. 
The straight line in the right figure is the result from the global fit.}
\label{fig:NEdn_full}
\vskip -5mm
\end{center}
\end{figure}

The top panels of Fig.~\ref{fig:NEdn_full} plot $R_3(E,t;\theta)$ 
as a function of $t$ for $E$ and $-E$ for each $K_{\rm val}$.
We clearly observe non-zero signals, which shows the expected dependence
on the sign of $E$. Furthermore the sign of the slope (positive for 
positive $E$) agrees with that in the quenched case.
By fitting $R_3^{\rm corr}$ with the exponential form (\ref{eq:R3corr})
over $6\le t-t_{\rm src}+1\le 10$ we obtain 
\begin{equation}
  d_N = -0.028(17)\,{\rm e \cdot fm},
\end{equation}
at $K=0.1357$. 
Similarly we have
\begin{equation}
  d_N = -0.033(22)\,{\rm e \cdot fm},\quad
  d_N = -0.032(25)\,{\rm e \cdot fm},
\end{equation}
at $K=0.1367$ and $K=0.1374$. 
The statistical error is still large.

\subsection{Quark mass dependence in quenched and full QCD}

As is well known, NEDM vanishes in the zero quark mass limit since 
$\theta$ term can be transformed to  $m\bar\psi e^{i\gamma_5\theta}\psi$ by
a chiral rotation. 
In quenched QCD, however, this property does not hold since the quark 
determinant is set to unity.  There exist no QCD-based theoretical
prediction for the quark mass dependence of NEDM in quenched QCD. 
The instanton liquid model \cite{Faccioli} predicts a 
$1/m^2_q \sim 1/m_\pi^4$ divergence, 
while partially quenched chiral perturbation 
theory\cite{Connell} leads to the $1/m_\pi^3$ behavior toward the chiral limit.

\begin{figure}
\begin{center}
\vskip -3mm
\includegraphics[width=60mm, angle=0] {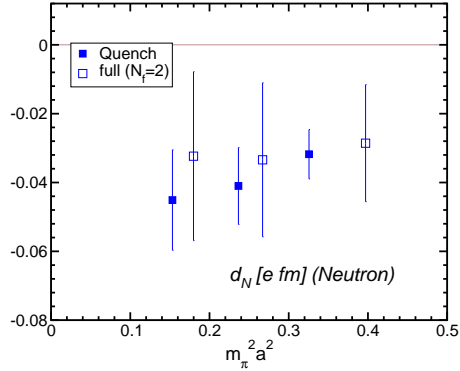}
\vskip -2mm
\caption{NEDM as a function of pion mass squared in quenched and full QCD.}
\label{fig:NEdn_mdep}
\vskip -5mm
\end{center}
\end{figure}

In Fig.~\ref{fig:NEdn_mdep} we plot values of NEDM as a function of pion
mass squared.
In quenched QCD these values are obtained after averaging over three
direction of the electric field. 
We take $8\le t-t_{\rm src}+1\le 12$ as the fitting range since
$t_{\rm src}=t_E=1$ is employed in this quenched calculation. 
We clearly observe a non-vanishing behavior toward the chiral limit in 
quenched QCD.
We are not able to determine, however, the precise
quark mass dependence to distinguish $1/m_\pi^4$ from $1/m_\pi^3$
due to large statistical errors and lack of sufficient quark mass points. 
In full QCD statistical errors are too large to confirm that NEDM
vanishes in the chiral limit.
We are currently increasing the number of
source points to reduce statistical errors. 

\section{Conclusions}

In this report we presented our lattice calculation of NEDM using 
the external electric field method. 
In quenched QCD we investigate the effect from the gap (discontinuity) of the electric
field, by comparing different two source points at the fixed gap point. 
From this investigation we learn that the effect of the gap to NEDM
signals can be reduced by placing the source point away from the gap.
We also obtained results which show that NDEM does not vanish toward the chiral limit, 
as indicated by some models.  This, however, is a quenching artifact. 

In two-flavor full QCD, we obtained, for the first time in 
full QCD simulations, a non-zero value for NEDM.  This is made possible 
by carefully avoiding the effect of the gap in the electric field by placing 
the source as far away as possible from the gap, and by accumulating 
a large number of statistics (over 12,000) by averaging over 8 source 
points and over both forward and backward propagation of the nucleon. 
The statistical error is still large (60-80\%).
Clearly we have to increase statistics in our full QCD calculations,
by employing more source points and/or different directions of the
electric field, in order to perform a reliable chiral extrapolation of
NEDM to the physical quark mass.
After this attempt, we will proceed to our final project, a 
calculation of NEDM in $N_f=2+1$ full QCD.

This work is supported in part by Grant-in-Aid of ministry of Education (Nos. 
13135204, 
13135216, 
15540251, 
16540228, 
17340066, 
17540259, 
18104005, 
18540250  
)
Numerical simulations are performed on Hitachi SR11000
at High Energy Accelerator Research
Organization (KEK), Hiroshima University and Tokyo University.
At KEK this simulation is under support of Large Scale
Simulation Program (No. 06-04).

\end{document}